\documentclass[aps,english,prl,floatfix,amsmath,superscriptaddress,tightenlines,twocolumn,nofootinbib]{revtex4-2}
\usepackage[left=1.5cm,right=1.5cm,top=2cm,bottom=2cm]{geometry}
\usepackage{xcolor}
\usepackage{graphicx,amsmath,wasysym}% Include figure files
\usepackage{dcolumn}% Align table columns on decimal point

\usepackage{bm}% bold math

\usepackage[colorlinks=true,linkcolor=blue,allcolors=blue]{hyperref}%
\usepackage{natbib}
\usepackage{geometry} 
\usepackage{ifthen}
\usepackage{slashed}
\usepackage{graphicx}
\usepackage{amssymb}
\usepackage{physics}
\usepackage{MnSymbol}
\usepackage{dsfont}
\usepackage{upgreek}
\usepackage{hyperref}
\usepackage{booktabs}       % For clean horizontal lines
\usepackage{array}
\usepackage{hhline}
\usepackage{bbold}

\usepackage[all,matrix,arrow,curve,arc]{xy}
\newcommand{\phdot}{{\phantom{\mkern1.3mu\cdot\mkern1.3mu}}}

\usepackage{tikz,xcolor,hyperref}

\definecolor{lime}{HTML}{A6CE39}
\DeclareRobustCommand{\orcidicon}{\hspace{-4pt}
\begin{tikzpicture}
\draw[lime, fill=lime] (0,0)
circle [radius=0.16]
node[white] {\hspace{0.1mm}{\fontfamily{qag}\selectfont \tiny ID}};
\draw[white, fill=white] (-0.07,0.1)
circle [radius=0.01];
\end{tikzpicture}
\hspace{-3.2mm}
}

\foreach \x in {A, ..., Z}{\expandafter\xdef\csname
orcid\x\endcsname{\noexpand\href{https://orcid.org/\csname orcidauthor\x\endcsname}
{\noexpand\orcidicon}}
}

\usepackage{pdfpages} % include pdfs
\usepackage{pgffor} % for loops
\makeatletter
\AtBeginDocument{\let\LS@rot\@undefined}
\makeatother

\usepackage{xr}
\makeatletter
\newcommand*{\addFileDependency}[1]{
 \typeout{(#1)}
 \@addtofilelist{#1}
  \IfFileExists{#1}{}{\typeout{No file #1.}}
}
\makeatother

\newcommand*{\myexternaldocument}[1]{
    \externaldocument{#1}
    \addFileDependency{#1.tex}
    \addFileDependency{#1.aux}
}
\myexternaldocument{Supplementary}
\usepackage{pdfpages}

\makeatletter
\AtBeginDocument{\let\LS@rot\@undefined}
\makeatother

\begin{document}
\title{Identifying chiral topological order in microscopic spin models by modular commutator}% Force line breaks with \\

\author{{Avijit Maity}\orcidA{}}
\email{avijit.maity@tifr.res.in}
\affiliation{Department of Theoretical Physics, Tata Institute of
Fundamental Research, Homi Bhabha Road, Colaba, Mumbai 400005, India}

\author{{Aman Kumar}\orcidB{}}
\email{akumar@magnet.fsu.edu}
\affiliation{National High Magnetic Field Laboratory, Tallahassee, Florida 32310, USA}
\affiliation{Department of Physics, Florida State University, Tallahassee, Florida 32306, USA}

\author{{Vikram Tripathi}\orcidC{}}
\email{vtripathi@theory.tifr.res.in}
\affiliation{Department of Theoretical Physics, Tata Institute of
Fundamental Research, Homi Bhabha Road, Colaba, Mumbai 400005, India}

%\date{\today}% It is always \today, today,
             %  but any date may be explicitly specified

\begin{abstract}
The chiral central charge $c_-$ is a key topological invariant of the edge characterizing the bulk two-dimensional chiral topological order, but its direct evaluation in microscopic spin models has long been a challenge, especially for non-abelian topological order. Building on the recently developed modular commutator formalism, we numerically obtain $c_-$ directly from single ground-state wave functions of two-dimensional interacting spin models that have chiral topological order. This provides a geometry-independent and bulk  diagnostic of chirality. We study two nonintegrable systems -- the Zeeman-Kitaev honeycomb model and the kagome antiferromagnet -- both subjected to scalar spin chirality perturbations. We find that the modular commutator yields results consistent with the expected topological quantum field theories. We also compute the topological entanglement entropy which provides an independent diagnostic of the topological orders. Our work establishes modular commutators as a powerful numerical probe of chiral topological order in strongly correlated quantum magnets.
\end{abstract}

\maketitle

\emph{Introduction.---} 
Topologically ordered phases in two dimensions are examples of long-range entangled (LRE) \cite{Kitaev_TEE_PRL2006, Levin_TEE_PRL2006} quantum states that cannot be characterized by conventional local order parameters and do not arise from spontaneous symmetry breaking \cite{Wen_VacuumDeg_PRB1989, Wen_FQHDeg_PRB1990,wen1990topological}. Prototypical instances of such phases are fractional quantum Hall states \cite{Laughlin_QHE_PRB1983} and quantum spin liquid (QSL) \cite{Savary_QSLReview_2016}, which host exotic quasiparticle excitations with fractional charge and anyonic statistics \cite{Leinaas:1977fm, Wilczek_Fractional_PRL1982, Arovas_Fractional_PRL1984}.

%\emph{Introduction.---} 
Chiral topological phases have gapless boundary modes (when defined on a manifold with edges) \cite{Wen:1990se, Wen_Gapless_PRB1991}, and for temperatures $T$ well below the bulk gap, the thermal Hall conductance $\kappa_{xy}$ is quantized in units of $(\pi k_B^2 /6\hbar)T$ \cite{Kane_Fisher_1997, Cappelli_2002}. The prefactor ($c_-$) is the \textit{chiral central charge}, which is a key topological invariant that characterizes the underlying gapped quantum phase.

More broadly, the universal data of a two-dimensional gapped phase without symmetries are captured by a unitary modular tensor category (UMTC)~\cite{Kitaev_Anyon_2006, Barkeshli_fractionalization_PRB2019, Rowell:2007dge, Bonderson:2007ci}, which encodes the fusion and braiding rules of anyonic excitations. Gauge-invariant quantities derived from the UMTC correspond to physical topological invariants such as the ground-state degeneracy, the modular $S$ and $T$ matrices, and the topological entanglement entropy (TEE) $\gamma$. A crucial exception is the chiral central charge $c_-$, which is not contained in the UMTC data but instead characterizes the gravitational anomaly of the associated edge conformal field theory (CFT). Two phases may share identical anyon statistics (same $\mathcal{C}$) yet differ in $c_-$; together, the pair $(\mathcal{C}, c_-)$ is believed to fully classify two-dimensional gapped phases without symmetries.

\begin{figure}[tb]
    \centering
    \begin{tikzpicture}[scale=0.36]
        % --- Outer boundary of the system ---
        \draw[] (-8,-4) -- (8,-4) -- (8,4) -- (-8,4) -- cycle;
        \node[below right] at (-7.8,3.8) {$\Lambda$};

        % --- Disk outline ---
        \draw[thick] (0,0) circle (2.6);

        % --- Colored sectors ---
        % Sector A (orange): from -90° to 30°
        \filldraw[fill=orange!50,draw=none,opacity=0.6]
            (0,0) -- (-90:2.6) arc (-90:30:2.6) -- cycle;

        % Sector B (blue): from 30° to 150°
        \filldraw[fill=blue!50,draw=none,opacity=0.6]
            (0,0) -- (30:2.6) arc (30:150:2.6) -- cycle;

        % Sector C (green): from 150° to 270°
        \filldraw[fill=green!50,draw=none,opacity=0.6]
            (0,0) -- (150:2.6) arc (150:270:2.6) -- cycle;

        % --- Sector dividers ---
        \draw[thick] (0,0) -- (30:2.6);
        \draw[thick] (0,0) -- (150:2.6);
        \draw[thick] (0,0) -- (-90:2.6);

        % --- Labels ---
        \node at (90:1.3) {\large $B$};
        \node at (-150:1.3) {\large $C$};
        \node at (-30:1.3) {\large $A$};
    \end{tikzpicture}

    \caption{Partition of a disk-shaped region $ABC$ in the bulk $\Lambda$ into three adjacent subsystems $A$, $B$, and $C$. This setup is used to implement the formulas for the chiral central charge and the TEE. Each subsystem is assumed to be sufficiently large compared to the correlation length so that universal contributions can be reliably extracted.}
    \label{fig:ABC_subsys}
\end{figure}
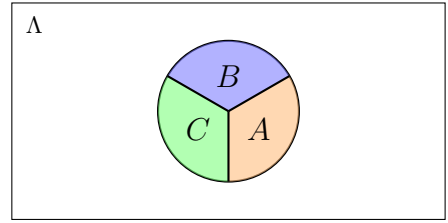

Although $c_-$ is expected to be a bulk property of the ground state, previous determinations have relied on a variety of indirect approaches, including momentum polarization \cite{Tu_Momentum_PRB2013}, gauge-theory analysis \cite{Guo_GaugeThermalHall_PRB2020,Maity_finiteTkxy_PRB2025,jagannath_gauge2025}, thermal Hall transport \cite{Kitaev_Anyon_2006, Kapustin_kxy_PRB2020,Guo_GaugeThermalHall_PRB2020,aman_thermal_hall2023, Maity_finiteTkxy_PRB2025}, and modular-matrix reconstruction from multiple minimum-entangled states (MES) typically obtained using infinite density-matrix renormalization group (iDMRG) \cite{Zhang_braiding_PRB2012,Cincio_Topological_PRL2013}. Each of these, however, relies on access to edge excitations, finite-temperature response, or the identification of several topological sectors. The methods for non-Abelian phases are even more limited \cite{Tu_Momentum_PRB2013, aman_thermal_hall2023}. These constraints make it challenging to evaluate $c_-$ in interacting systems solely using the microscopic ground-state wavefunctions. To overcome these limitations, Refs.~\cite{Kim_c_minus_PRL2022, Kim_Modular_PRB2022} recently obtained a bulk formula for $c_-$ that requires only a \textit{single} ground state wave function, expressed in terms of the commutator of modular Hamiltonians obtained from that wave function. For a general tripartite state $\rho_{ABC}$ defined on a finite-dimensional Hilbert space, the modular commutator takes the form \cite{Kim_c_minus_PRL2022, Kim_Modular_PRB2022}
\begin{align}
\label{eq:modular_commutator}
    J(A,B,C) = i \Tr \left(\rho_{ABC}\left[K_{AB},K_{BC}\right]\right),
\end{align}
where  $K_X = - \ln \rho_X $ is the modular Hamiltonian of the reduced density matrix $\rho_X$ in region $X$. 

For three adjoining subsystems $A$, $B$, and $C$ of a many-body ground state $\ket{\psi}$ that satisfies the area law with a constant subleading term (see Fig.~\ref{fig:ABC_subsys}) \cite{Kitaev_TEE_PRL2006, Levin_TEE_PRL2006}, the modular commutator is directly related to the chiral central charge \cite{Kim_c_minus_PRL2022, Kim_Modular_PRB2022, Fan_real-space_SciPost2023}:
\begin{align}
\label{eq:c_minus}
    J(A,B,C) = \frac{\pi}{3} c_-.
\end{align}

This relation in Eq.~\eqref{eq:c_minus} has been numerically verified in non-interacting models, including the Chern insulator and the $p+ip$ topological superconductor \cite{Zou_Commutators_CFT_PRL2022, Fan_Gravitational_PRL2022, Kim_Virasoro_Annals2024, Park_additivity_PRB2025}. For interacting systems, Ref.~\cite{Kim_Modular_PRB2022} performed a test on the lattice realization of the $\nu=1/2$ bosonic Laughlin state \cite{Laughlin_Anomalous_PRL1983, Kalmeyer-Laughlin_RVB-FQH_PRL1987}, using a ground state wave function ansatz previously derived from CFT arguments \cite{Nielsen_Laughlin_CFT_PRL2012}, and confirmed the formula numerically up to small errors attributable to finite-size effects. However, direct validation of  chiral topological order in genuinely interacting microscopic spin Hamiltonians using the modular commutator method has not yet been attempted. This would be particularly useful to determine  $c_{-}$ for non-Abelian chiral phases such as Ising topological order (ITO) for which only limited techniques are available.

In this Letter, we bridge the aforementioned gaps by computing $c_-$ (and the TEE $\gamma$) directly from the ground-state wavefunctions of two paradigmatic chiral spin liquids (CSLs). Our analysis is based on exact diagonalization of finite clusters. As representative non-integrable, interacting spin models, we study the Zeeman-Kitaev honeycomb model with three-spin interactions and the kagome Heisenberg model with a scalar spin-chirality term. In the non-Abelian phase of the Kitaev model, we obtain $c_- = \frac{1}{2}$ and $\gamma = \ln 2$, consistent with its description by the Ising topological quantum field theory (TQFT) \cite{Kitaev_Anyon_2006}. For the kagome model, our results yield $c_- = 1$ and $\gamma = \frac{1}{2}\ln 2$, in agreement with the Abelian $U(1)_2$ Chern–Simons theory corresponding to the $\nu=1/2$ bosonic Laughlin state \cite{Bauer_KagomeCSL_Nature2014, Kumar-Fradkin_CSLKagome_PRB2015, Maity_finiteTkxy_PRB2025}.

\iffalse
In this Letter, we compute the chiral central charge for both non-Abelian and Abelian chiral topological orders, illustrated by the Kitaev honeycomb model and the kagome Heisenberg model, respectively. In the non-Abelian phase of the Kitaev model, we obtain $c_- = \frac{1}{2}$, consistent with its description by an Ising topological quantum field theory (TQFT) \cite{Kitaev_Anyon_2006}. For the kagome Heisenberg model with a scalar spin chirality term, we find $c_- = 1$, in agreement with the expected Abelian Chern–Simons theory at level $k=2$ \cite{Bauer_KagomeCSL_Nature2014, Kumar-Fradkin_CSLKagome_PRB2015, Maity_finiteTkxy_PRB2025}.
\fi

\begin{figure}
    \centering
    \includegraphics[width=\linewidth]{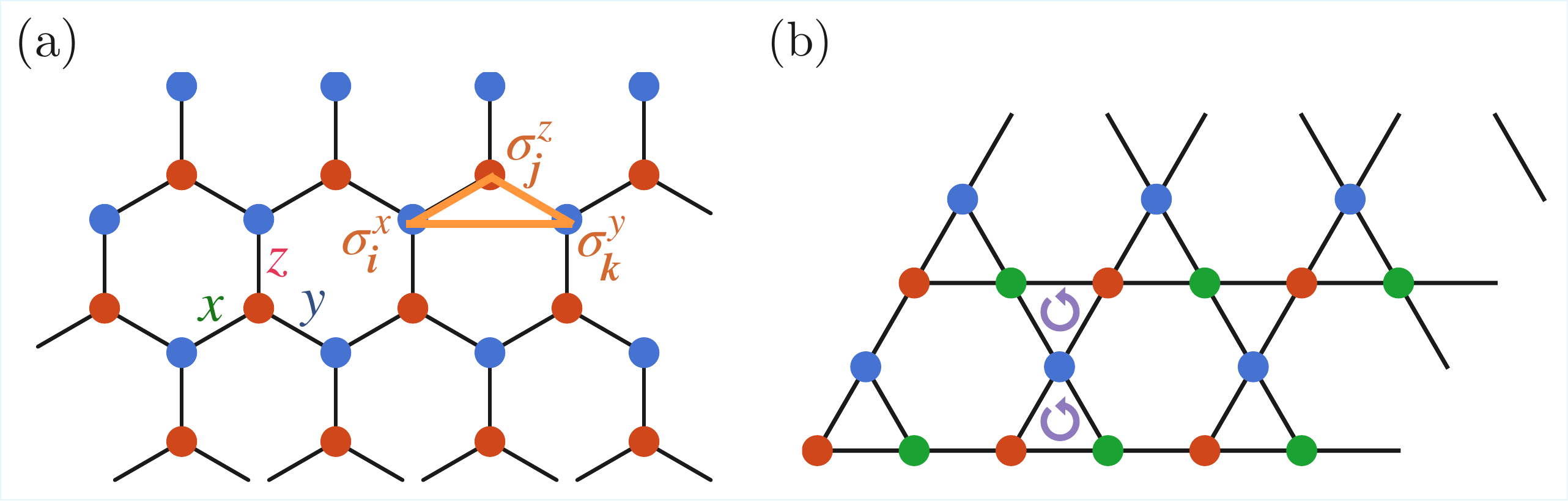}
    \caption{(\textbf{a}) Honeycomb lattice for the Kitaev model, with nearest-neighbor bonds labeled by $x$, $y$, and $z$, and a single three-spin term $\sigma^x_i \sigma^y_j \sigma^z_k$ of the three-spin chirality term; (\textbf{b}) Schematic of the kagome lattice. The sites $(i,j,k)$ on a triangular plaquette are ordered counterclockwise to define the scalar spin-chirality interaction.}
    \label{fig:Two_lattice}
\end{figure}

\emph{Kitaev honeycomb model.---} 
The Hamiltonian of the spin-$1/2$ Kitaev model with an added three-spin interaction term is written as
\begin{align}
\label{eq:Kitaev_model}
    H = & \sum_{\langle i,j \rangle \in \mu\, \text{links}} K_\mu \sigma^\mu_i \sigma^\mu_j - K_3 \sum_{\llangle i,j,k \rrangle} \sigma^x_i \sigma^y_j \sigma^z_k \nonumber\\
    & - h \sum_i \left( \sigma^x_i + \sigma^y_i +\sigma^z_i \right),
\end{align}
where the first term corresponds to the pure Kitaev model, with $\mu = x, y, z$ labeling the three types of nearest-neighbor bonds $\langle i,j \rangle$ on the honeycomb lattice as shown in Fig.~\ref{fig:Two_lattice}a. The second term corresponds to the three-spin chirality term, which explicitly breaks time-reversal symmetry. The notation $\llangle \cdot \rrangle$ indicates an ordered tuple $(i,j,k)$ of neighboring sites, chosen such that the spin components $\sigma^x$, $\sigma^y$, and $\sigma^z$ at the two outer sites are aligned with the bond labels connecting them to the central site \cite{Gohlke_Dynamical_PRB2018} (see Fig.~\ref{fig:Two_lattice}a). The third term represents the Zeeman interaction of spins with a magnetic field applied along the $[1,1,1]$ direction, which destroys the exact integrability of the Kitaev model.

\iffalse
\begin{align}
    H = & \sum_{\langle i,j \rangle \in x\, \textit{links}} K_x \sigma^x_i \sigma^x_j + \sum_{\langle i,j \rangle \in y\, \textit{links}} K_y \sigma^y_i \sigma^y_j \nonumber\\
    & + \sum_{\langle i,j \rangle \in z\, \textit{links}} K_z \sigma^z_i \sigma^z_j + K_3 \sum_{\langle \langle i,j,k \rangle \rangle} \sigma^x_i \sigma^y_j \sigma^z_k.
\end{align}
\fi

At $K_3 = 0$ and $h=0$, the Kitaev model exhibits a QSL ground state characterized by fractionalized excitations. The fermionic spectrum is determined by $K_\mu$: it yields a gapped $\mathbb{Z}_2$ spin liquid with Abelian quasiparticles ($A$-phase) or remains gapless ($B$-phase) \cite{Kitaev_Anyon_2006}. The gapless phase can be driven into a non-Abelian topological phase—known as Ising topological order (ITO)—through time-reversal symmetry breaking perturbations, such as the inclusion of a three-spin chirality term \cite{Lee_Solitons_PRL2007} or the application of an external magnetic field \cite{Kitaev_Anyon_2006}.

In addition to analyzing the chiral central charge, we also evaluate the TEE as an independent check. The TEE can be extracted directly from the ground-state wave function of a topologically ordered phase via the linear combination of entanglement entropies \cite{Kitaev_TEE_PRL2006, Levin_TEE_PRL2006},
\begin{align}
\label{eq:TEE_gamma}
\gamma = S_{AB} + S_{BC} + S_{AC} - S_{A} - S_{B} - S_{C} - S_{ABC},
\end{align}
where $S_X$ is the von Neumann entropy of region $X$, defined by tracing over the degrees of freedom outside $X$, and $ABC$ is the tripartite bulk disk as shown in Fig.~\ref{fig:ABC_subsys}.

In the ITO (non-Abelian) phase, the Kitaev model realizes the Ising TQFT supporting anyons $\mathbb{1}$, $\sigma$, and $\psi$ \cite{Kitaev_Anyon_2006, SM}. The corresponding universal topological quantities are well known: $c_- = \tfrac{1}{2}$ and $\gamma = \ln 2$. The chiral central charge of the ITO phase of the Kitaev model has been obtained earlier using momentum polarization \cite{Tu_Momentum_PRB2013}, gauge theory approach \cite{jagannath_gauge2025}, and from direct numerical microscopic calculation of the edge thermal Hall current at finite temperature \cite{aman_thermal_hall2023}. The momentum polarization approach makes use of integrability, which is absent in the presence of integrability breaking perturbations such as a Zeeman term we have here. The gauge theory approach requires a proper mean field solution around which gauge fluctuations are considered, and the finite temperature edge current calculation requires the use of excited state properties. Here we compute $c_{-}$ using the modular commutator approach \eqref{eq:c_minus}, which provides a bulk, geometry-independent diagnostic from a single ground-state wavefunction.

To benchmark our numerical approach, we perform exact diagonalization on finite clusters of size $N = L_x \times L_y \times 2$, with $L_x$ and $L_y$ denoting the number of unit cells along the $x$ and $y$ directions, respectively. We study clusters of size $N \in {12,16,18,24,30}$, all having periodic boundary conditions and extract the ground state in the $(0,0)$ momentum sector. The finite cluster geometries employed in our study and the subsystem partitions used to evaluate $c_-$ and $\gamma$ are shown in the Supplemental Material \cite{SM}. To remain deep in the ITO phase, we choose isotropic couplings $K_x = K_y = K_z = 1$ and set $h=0.05,\, K_3 = 0.4$ \cite{Fu_Kitaev_PRB2018}.

\begin{figure}[tb]
    \centering
    \includegraphics[width=\linewidth]{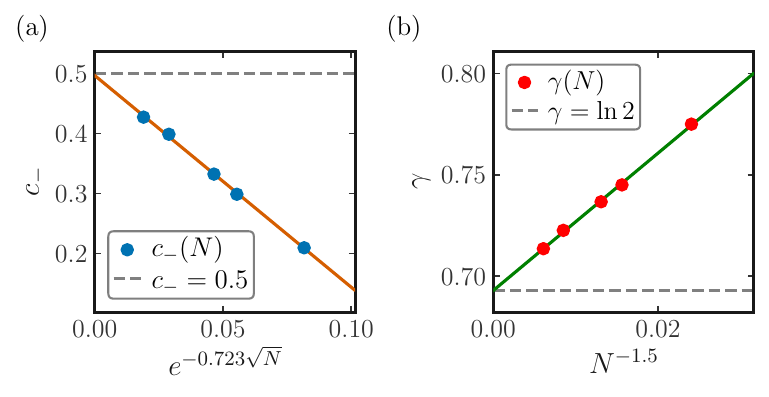}
    \caption{Finite-size scaling of (\textbf{a}) the chiral central charge $c_-$ and (\textbf{b}) the TEE $\gamma$ in the ITO phase of the Kitaev model. The numerical results (dots) are obtained from exact diagonalization of finite clusters, and the solid curves show the best fits using exponential (for $c_-$) and power-law (for $\gamma$) ansatz functions. The horizontal dashed lines indicate the theoretical values, $c_- = \frac{1}{2}$ and $\gamma = \ln 2$, respectively.}
    \label{fig:Kitaev_c_minus_gamma}
\end{figure}

For the largest cluster studied ($N=30$), we obtain $c_- \approx 0.427$ and $\gamma \approx 0.714$, corresponding to deviations of $14.4 \%$ and $3.0 \%$, respectively, from the theoretical predictions. These deviations systematically decrease with increasing system size, as illustrated in Fig.~\ref{fig:Kitaev_c_minus_gamma}a. Such finite-size effects are expected and were also observed in Ref.~\cite{Kim_Modular_PRB2022}. To quantify the convergence toward the thermodynamic limit, we follow the fitting procedure of Ref.~\cite{Kim_Modular_PRB2022}, employing two families of ansatz functions,
\begin{align}
& f_p(x) = a x^{-b} + c, \nonumber \\
& f_e(x) = a e^{-b x^d} + c,
\end{align}
which capture ``power-law'' and ``exponential'' finite-size scaling, respectively. The quality of the fit is judged solely by its deviation from the numerical data points, with theoretical values used only for reference in Figs.~\ref{fig:Kitaev_c_minus_gamma}a and \ref{fig:Kitaev_c_minus_gamma}b.

For the chiral central charge $c_-$, we find that the exponential ansatz provides a clearly superior description of the data, consistent with the conclusions of Ref.~\cite{Kim_Modular_PRB2022}. The best fit is given by
\begin{align}
c_-(N) = -3.5274 e^{-0.723 \sqrt{N}} + 0.4981,
\end{align}
from which we extract the extrapolated value
\begin{align}
c_-(\infty) = 0.4963 \pm 0.0159.
\end{align}
This result is in excellent agreement with the theoretical value $c_- = \tfrac{1}{2}$, with a relative error of only $0.7 \%$ in the thermodynamic limit.

In contrast, the TEE data are best captured by the power-law ansatz, with an exponent $b = 1.5$ (see Fig.~\ref{fig:Kitaev_c_minus_gamma}b). The optimal fit,
\begin{align}
\gamma(N) = 3.387 N^{-1.5} + 0.693,
\end{align}
yields the extrapolated value
\begin{align}
\gamma(\infty) = 0.693 \pm 0.0009,
\end{align}
which reproduces the expected $\ln 2$ with remarkable precision, corresponding to a relative error of only $0.02 \%$.

Together, these results demonstrate that both the chiral central charge and the TEE extracted from exact diagonalization converge rapidly to their universal values in the thermodynamic limit, thereby providing strong numerical evidence for the realization of ITO in this regime of the Kitaev model.

\emph{Kagome chiral spin liquid.---}
Next, we study a spin-$1/2$ antiferromagnet on the kagome lattice described by the Hamiltonian
\begin{equation}
\label{eq:H_kagome}
    H = J_1 \sum_{\langle \bm{i j} \rangle} \bm{S}_{\bm{i}} \cdot \bm{S}_{\bm{j}} + J_\chi \sum_{\bm{i j k} \in \Delta, \nabla} \bm{S}_{\bm{i}} \cdot (\bm{S}_{\bm{j}} \times \bm{S}_{\bm{k}}),
\end{equation}
where $\bm{S}_{\bm{i}}$ denotes the spin-$1/2$ operator at site $\bm{i}$. The first term represents the nearest-neighbor Heisenberg interaction, which respects global SU$(2)$ spin-rotation symmetry, time-reversal symmetry, and all discrete lattice symmetries of the kagome lattice. The second term is the scalar spin-chirality interaction, where the sum runs over all triangular plaquettes with sites $(i,j,k)$ ordered counterclockwise as shown in Fig.~\ref{fig:Two_lattice}b. This term breaks time-reversal and parity symmetries while preserving global SU$(2)$ invariance. The coupling $J_\chi$ may arise, for example, from an external magnetic field, in which case its magnitude is proportional to the sine of the magnetic flux through each elementary triangle of the kagome lattice \cite{Diptiman_Chiral_PRB1995}.

In kagome antiferromagnets, a CSL can be stabilized by scalar spin-chirality interactions \cite{Bauer_KagomeCSL_Nature2014}. Using the iDMRG, Ref.~\cite{Bauer_KagomeCSL_Nature2014} showed that sufficiently strong time-reversal symmetry breaking drives the ground state into a CSL belonging to the same topological class (semionic order) as the $\nu=1/2$ bosonic Laughlin state \cite{Kalmeyer-Laughlin_RVB-FQH_PRL1987, Kalmeyer-Laughlin_SL-Heisenberg_PRB1989}. CSLs may also arise spontaneously in kagome Heisenberg antiferromagnets with further-neighbor interactions \cite{Gong-Sheng_CSL_Z2_Nature2014}, and have been reported in a variety of other theoretical studies across different models \cite{Yin-Chen-He_CSL_Kagome_PRL2014, Yin-Chen-He_XXZ_Kagome_PRL2015, Sheng_VMC_CSL_kagome_PRB2015, Gong-Balents-Sheng_GlobalPhase_PRB2015, Bieri-Lhuillier_Gapless_CSL_PRB2015, Yin-Subhro-Pollmann_PRL2015, Kumar-Fradkin_CSLKagome_PRB2015, Niu_KagomeCSl_iPess, Maity_finiteTkxy_PRB2025}.

The $\nu = 1/2$ bosonic Laughlin state is captured by the Abelian $U(1)_2$ Chern–Simons theory \cite{Bauer_KagomeCSL_Nature2014, Kumar-Fradkin_CSLKagome_PRB2015, Maity_finiteTkxy_PRB2025}, which supports a single nontrivial anyon, the semion $s$ \cite{SM}. The universal topological invariants are well known: the chiral central charge is $c_- = 1$, while the topological entanglement entropy is given by $\gamma = \frac{1}{2}\ln 2$ \cite{Maity_finiteTkxy_PRB2025}.

The chiral central charge $c_{-}$ of the kagome Heisenberg model with scalar spin chirality in Eq.~\eqref{eq:H_kagome} has been obtained from several complementary approaches: from edge CFT and bulk thermal Hall transport, both based on parton mean-field theory~\cite{Maity_finiteTkxy_PRB2025}, and from modular-matrix analysis using iDMRG~\cite{Bauer_KagomeCSL_Nature2014}. In the latter, Bauer et al. employed the framework of Refs.~\cite{Zhang_braiding_PRB2012,Cincio_Topological_PRL2013}, constructing torus wavefunctions from MES on an infinite cylinder and extracting the modular $S$ and $T$ matrices from their overlaps under a $\pi/3$ rotation. The overall phase of the $T$ matrix, $T_{\mathbb{1} \mathbb{1}} \propto e^{-i2\pi c_{-}/24} $, directly encodes the chiral central charge. The resulting matrices matched those of the semion ($\nu=1/2$ bosonic Laughlin) topological order, giving $c_{-}\simeq1$, consistent with a single chiral boson edge mode.

However, subsequent analyses have shown that extracting the full modular data from MES is a subtle process: the correspondence between MES and anyon types is often ambiguous, MES overlaps alone generally provide only partial information (such as fusion rules and composite phase factors), and additional lattice symmetries are required to fully fix the $S$ and $T$ matrices~\cite{Zhang_braiding_PRB2012}.
In contrast, we determine $c_{-}$ using the modular-commutator approach~\eqref{eq:c_minus}, which offers a purely bulk, lattice-symmetry independent wavefunction-based diagnostic that requires only a single ground state~\cite{Kim_c_minus_PRL2022}.

\begin{figure}
    \centering
    \includegraphics[width=\linewidth]{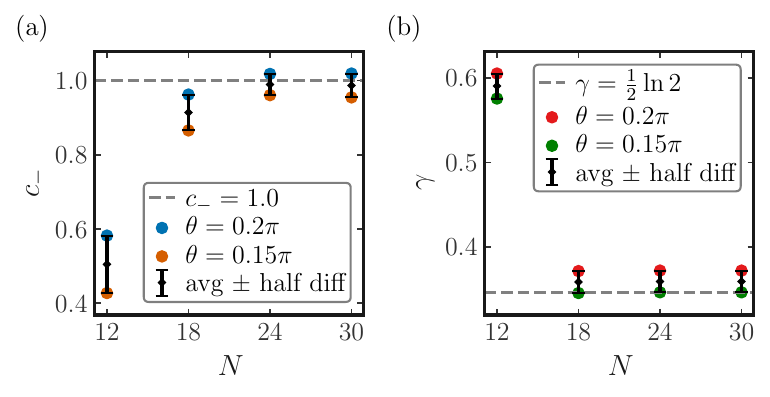}
    \caption{ Finite-size results for (\textbf{a}) the chiral central charge $c_-(N)$ and (\textbf{a}) the TEE $\gamma(N)$ of the kagome CSL, obtained from exact diagonalization on clusters with $N=12,18,24,30$ sites and periodic boundary conditions. Blue (orange) symbols denote raw data at $\theta=0.2\pi$ ($\theta=0.15\pi$). Black diamonds show the pairwise averages of the two datasets at each $N$, with error bars corresponding to half the inter-$\theta$ difference and thus providing a conservative estimate of systematic uncertainty within the CSL phase. Dashed horizontal lines indicate the universal values expected for the $\nu = 1/2$ bosonic Laughlin state, $c_- = 1$ and $\gamma = \tfrac{1}{2}\ln 2$. The near-convergence of the averaged data for $N=24$ and $N=30$ demonstrates consistency with the universal topological invariants.}
    \label{fig:Kagome_c_minus_gamma}
\end{figure}

We analyze the kagome Hamiltonian \eqref{eq:H_kagome} using exact diagonalization on finite clusters of size $N = L_x \times L_y \times 3$, with $L_x$ and $L_y$ denoting the number of unit cells along the $x$ and $y$ directions, respectively. We study clusters of size $N \in {12,18,24,30}$, all with periodic boundary conditions. Details of the finite cluster geometries and the subsystem partitions for $c_-$ and $\gamma$ are presented in the Supplementary Material \cite{SM}. Since the Hamiltonian commutes with the uniform magnetization operator $\bm{M}$, we work in the $M^z = \sum_{\bm{i}} S^z_{\bm{i}}$ basis and restrict to the ground-state sector with minimum magnetization, i.e., $M^z=0$ for even-spin clusters.

For convenience, we parametrize the couplings as $J_{1} = J \cos\theta$ and $J_\chi = J \sin\theta$, with $J=1$. It is known from Ref.~\cite{Bauer_KagomeCSL_Nature2014} that the CSL phase is remarkably robust: the semionic topological order persists almost up to the pure Heisenberg point, and the CSL remains stable for all $\theta \gtrsim 0.05\pi$. This broad stability window ensures that both $\theta=0.15\pi$ and $\theta=0.2\pi$ lie well within the CSL region. Motivated by this, we compute the topological invariants at both parameter values. The two datasets show the same qualitative finite-size behavior but differ slightly at small $N$ due to non-monotonic size effects as shown in Fig.~\ref{fig:Kagome_c_minus_gamma}. To reduce such fluctuations, at each cluster size we take the average of the two values and use half their difference as a conservative estimate of the systematic uncertainty associated with varying $\theta$ inside the CSL phase. The resulting finite-size behavior of the chiral central charge $c_-(N)$ and TEE $\gamma(N)$ is summarized in Fig.~\ref{fig:Kagome_c_minus_gamma}.

For the chiral central charge we find $c_-(N=12) \approx 0.505 \pm 0.077$, $c_-(N=18) \approx 0.914 \pm 0.048$, $c_-(N=24) \approx 0.989 \pm 0.029$, and $c_-(N=30) \approx 0.987 \pm 0.032$. The relatively low value at $N=12$ reflects strong finite-size suppression, while the rapid approach toward unity by $N=18$–$30$ indicates that finite-size effects decay quickly. Crucially, the two largest clusters already agree within less than $0.5 \%$, with overlapping error bars, yielding a controlled large-$N$ estimate $c_- \approx 0.99$. In contrast to the Kitaev model, where we employed a scaling ansatz to extrapolate to $N \to \infty$, here the kagome data exhibit clear convergence without the need for such an extrapolation. We therefore present only the raw finite-size data in Fig.~\ref{fig:Kagome_c_minus_gamma}a, and the large-$N$ values are fully consistent with the universal expectation $c_-=1$.

Similarly, for the TEE we obtain $\gamma(N=12) \approx 0.590 \pm 0.015$, $\gamma(N=18) \approx 0.358 \pm 0.013$, $\gamma(N=24) \approx 0.359 \pm 0.013$, and $\gamma(N=30) \approx 0.359 \pm 0.013$. The smallest cluster strongly overshoots the asymptotic value, while $N=18$ and larger already cluster around the expected $\gamma = \tfrac{1}{2}\ln 2 \approx 0.347$. Again, the $N=24$ and $N=30$ results are essentially converged, so we simply report the raw data in Fig.~\ref{fig:Kagome_c_minus_gamma}b.

Taken together, the rapid finite-size convergence of $c_-(N)$ and $\gamma(N)$ at large $N$, together with the broad stability of the CSL phase, provides compelling numerical evidence that the kagome antiferromagnet with scalar spin chirality realizes the $\nu=1/2$ bosonic Laughlin state, i.e., a CSL with semionic topological order.

\emph{Discussion.---}
In summary, we have computed directly from microscopic spin models two universal topological quantities—the chiral central charge $c_-$ and the TEE $\gamma$—for both non-Abelian and Abelian chiral spin liquids. For the Kitaev model with three-spin interactions, we employed a finite-size scaling ansatz to extrapolate exact diagonalization results to the thermodynamic limit, obtaining $c_-(\infty) = 0.4963 \pm 0.0159$ and $\gamma(\infty) = 0.693 \pm 0.0009$, in excellent agreement with the predictions of Ising topological order. For the kagome antiferromagnet with scalar spin chirality, by contrast, the finite-size data show clear saturation without the need for extrapolation: the two largest clusters already yield $c_- \approx 0.99$ and $\gamma \approx 0.359$, consistent with the $\nu=1/2$ bosonic Laughlin state with semionic order.

Our results demonstrate that $c_-$ and $\gamma$ can be reliably extracted from computationally modest finite-size calculations. This provides a flexible and powerful numerical diagnostic of chiral topological order in strongly correlated quantum magnets. The same framework can be readily extended to other lattice models and may also prove useful in probing topological responses at finite temperature.

A.M. and V.T. acknowledge support of the Department
of Atomic Energy, Government of India, under Project
Identification No. RTI 4002, and the Department of Theoretical Physics, TIFR, for computational resources. A.K.
thanks the National High Magnetic Field Laboratory,
which is supported by National Science Foundation Cooperative Agreement No. DMR-2128556* and the State
of Florida. A.K. was supported through a Dirac postdoctoral fellowship at NHMFL. Exact diagonalization was performed using the QuSpin package \cite{QuSpin_SciPost2017, QuSpin_SciPost2019}.

\bibliography{bibliography}
\iffalse
\newpage
\foreach \x in {1,...,3}
{
\clearpage
\includepdf[pages={\x},angle=0]{Supplementary} 
}
\fi

\onecolumngrid

\newpage

\setcounter{equation}{0}
\renewcommand{\theequation}{SM\arabic{equation}}
\setcounter{section}{0}

\begin{center}
{\bf Supplementary Material for ``Identifying chiral topological order in microscopic spin models by modular commutator''}
\end{center}

%\tableofcontents

\section*{Unitary modular tensor category theory and chiral central charge}

The algebraic theory of anyons provides a unified framework for describing two-dimensional topological order. It classifies gapped phases through the fusion and braiding of quasiparticles, formalized by unitary modular tensor categories (UMTCs). While UMTCs offer a Hamiltonian-independent description, intuition often comes from exactly solvable models. We now briefly review the key UMTC data commonly used in the literature, following Refs.~\cite{Kitaev_Anyon_2006, Barkeshli_fractionalization_PRB2019, Rowell:2007dge, Bonderson:2007ci, Li_modular_PRB2022}.

In the language of topological quantum field theory (TQFT), the set of superselection sectors forms a UMTC, denoted $\mathcal{C}$. Its elements, called anyonic charges $a,b,c,\dots \in \mathcal{C}$, encode the fusion and braiding properties of anyons.

The fusion rules take the form
\begin{align}
    a \times b = \sum_{c \in \mathcal{C}} N_{ab}^c c,
\end{align}
where $N_{ab}^c$ are non-negative integers specifying how many distinct ways charges $a$ and $b$ can fuse into $c$. From this framework, one obtains the key modular data of the TQFT—quantum dimensions $d_a$, topological spins $\theta_a$, total quantum dimension $D$, and the modular matrices $S$ and $T$—as summarized in Table~\ref{tab:UMTC}.

\begin{table}[h]
\caption{Modular data of a TQFT: anyon labels $a$, quantum dimensions $d_a$, topological spins $\theta_a$, total quantum dimension $\mathcal{D}$, and modular matrices $S$ and $T$.
}
\label{tab:UMTC}
\begin{ruledtabular}
\begin{tabular}{l p{0.78\columnwidth}}
\textbf{Notation} & \textbf{Definition} \\ \hline
$\mathcal{C}$ & A UMTC (TQFT model); it encodes the set of anyons along with their fusion and braiding properties. \\
$a,b,c,\ldots$ & The anyons in $\mathcal{C}$ are characterized by fusion rules that are both commutative and associative.\\
$N^{\,c}_{ab}$ & They are non-negative integers that indicate the number of distinct ways the charges $a$ and $b$ can fuse into $c$. \\
$d_a$ & The quantum dimension of the anyon $a$, $d_a = {\xy (-5,0)="a";  "a";"a"+(0,1),**\dir{}, "a",{\ellipse(5){-}}; {\ar@{>} (-0.05,0.85);(-0.05,0.9)}; (2,0)*{ a} \endxy} $. An anyon is non-Abelian exactly when its quantum dimension satisfies $d_a > 1$. \\
$\mathcal{D}$ & Total quantum dimension of an anyon model, $ \mathcal{D} = \sqrt{ \sum_{a\in \mathcal{C}} {d_a^2}}$.\\
$\theta_a$ & The topological twist (or topological spin) of anyon $a$, which can be derived from the braiding: $\theta_a = \frac{1}{d_a} {\xy (0,0)*{\phdot}="o"; (-3,3)="tl"; (3,3)="tr"; (-3,-3)="bl"; (3,-3)="br"; (3,5)*{a}; %"tl"+(0,1.5)*{\scr a}; 
			"o";"tl"**\dir{-}; "bl",{\ellipse^{}}; "tr"**\dir{-}?(1)*\dir{>}; "br",{\ellipse_{}}; "o"**\dir{-}; \endxy}$  \\
$S$ & The modular $S$ matrix describes the mutual statistics of anyons: $\displaystyle S_{a,b} =  \mathcal{D}^{-1} \sum_c N_{ab}^c \frac{\theta_c}{\theta_a\theta_b} d_c=\frac{1}{\mathcal{D}}{\xy (-9,0)*{}="Bl"; (-4,0)="a"; (2,0)="b"; (1,0.7)="aa",*+!L{a}; {\ar@{>} "aa"+(0,-.01);"aa"}; (7,0.7)="bb",*+!L{ b}; {\ar@{>} "bb"+(0,-.01);"bb"};
		"a";"b",**\dir{}, "a",{\ellipse(5):a(154),=:a(-20){-}}; "b";"a",**\dir{}, "b",{\ellipse(5):a(154),=:a(-20){-}}; \endxy}.$  \\
$T$ & The modular $T$ matrix describes the self statistics of anyons: $T_{a,b} = \delta_{a,b}\theta_{a}$.  \\
\end{tabular}
\end{ruledtabular}
\end{table}

The chiral central charge $c_-$ is not part of the UMTC itself. However, for bosonic topologically ordered systems, there is a remarkable formula that relates $c_-$ modulo $8$ to the bulk anyon data \cite{Kitaev_Anyon_2006}:
\begin{align}
    \mathcal{D}^{-1} \sum_a d_a^2 \, \theta_a = e^{2\pi i c_-/8}. 
\end{align}
It is also widely conjectured that the combination of a UMTC and $c_-$ completely classifies all two-dimensional gapped phases without symmetries.

Here, we present two examples of anyon data with their chiral central charge—the Ising UMTC and the Semion UMTC—both of which were studied in the main text.

\paragraph{Ising UMTC:} The anyon data and chiral central charge for the Ising anyon model \cite{Kitaev_Anyon_2006, Rowell:2007dge}:
\begin{itemize}
    \item Superselection sectors: $\mathcal{C} = \{\mathbb{1},\sigma,\psi\}$.
    \item Fusion rules: $\sigma \times \sigma = \mathbb{1}+\psi$, $\sigma \times \psi = \sigma$, $\psi \times \psi = \mathbb{1}$
    \item Quantum dimensions: $d_{\mathbb{1}} = 1$, $d_\sigma=\sqrt{2}$, $d_\psi=1$.
    \item Total quantum dimension: $\mathcal{D} = 2$.
    \item Topological $S$-matrix:
    \begin{align}
        S = \frac{1}{2} \begin{pmatrix}
1 & \sqrt{2} & 1 \\
\sqrt{2} & 0 & -\sqrt{2} \\
1 & - \sqrt{2} & 1
\end{pmatrix}.
    \end{align}
    \item  Topological spins: $\theta_{\mathbb{1}} =1$, $\theta_\sigma = e^{\frac{i\pi}{8}} $, $\theta_\psi = -1$.
    \item  Chiral central charge: $ c_- = \frac{1}{2}$. 
\end{itemize}

\paragraph{Semion UMTC:} The anyon data and chiral central charge for the semion anyon model \cite{Rowell:2007dge}:
\begin{itemize}
    \item Superselection sectors: $\mathcal{C} = \{\mathbb{1},s\}$.
    \item Fusion rules: $s \times s = \mathbb{1}$,
    \item Quantum dimensions: $d_{\mathbb{1}} = 1$, $d_s=1$.
    \item Total quantum dimension: $\mathcal{D} = \sqrt{2}$.
    \item Topological $S$-matrix:
    \begin{align}
        S = \frac{1}{\sqrt{2}} \begin{pmatrix}
1 & 1 \\
1 & -1
\end{pmatrix}.
    \end{align}
    \item  Topological spins: $\theta_{\mathbb{1}} =1$, $\theta_s = i$.
    \item  Chiral central charge: $ c_- = 1$.
\end{itemize}

%\fi
%\section*{Clusters for exact diagonalization under periodic boundary conditions}
\section*{Exact diagonalization clusters geometries}
\begin{figure}[h]
    \centering
    \includegraphics[width=\linewidth]{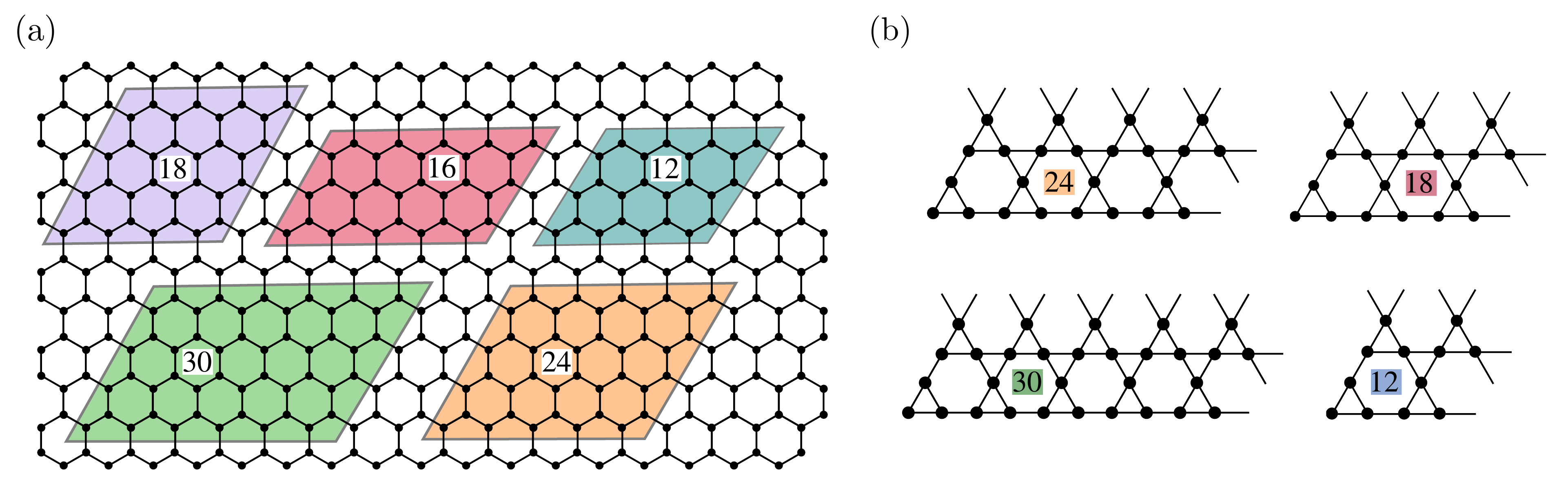}
    \caption{Finite clusters used for exact diagonalization. (\textbf{a}) Honeycomb clusters for the Kitaev model with $N=12,16,18,24,30$ sites. (\textbf{b}) Kagome clusters for the Heisenberg model with scalar spin-chirality, with $N=12,18,24,30$ sites. All clusters are taken with periodic boundary conditions.}
    \label{fig:Cluster_honeycomb_kagome}
\end{figure}

To carry out exact diagonalization (ED), we studied finite clusters with periodic boundary conditions for both the Kitaev honeycomb model and the kagome Heisenberg model with scalar spin-chirality interactions. The clusters were chosen to preserve as many lattice symmetries as possible, while allowing access to a range of system sizes. For the Kitaev model, we considered clusters with $N=12,16,18,24,30$ sites. For the kagome model, we used clusters with $N=12,18,24,30$ sites. The shapes of all clusters employed in our calculations are illustrated in Fig.~\ref{fig:Cluster_honeycomb_kagome}, where each cluster is shown with its periodic identifications.

\section*{Subsystem partitions for $c_-$ and $\gamma$}

\begin{figure}[h]
    \centering
    \includegraphics[width=0.85\linewidth]{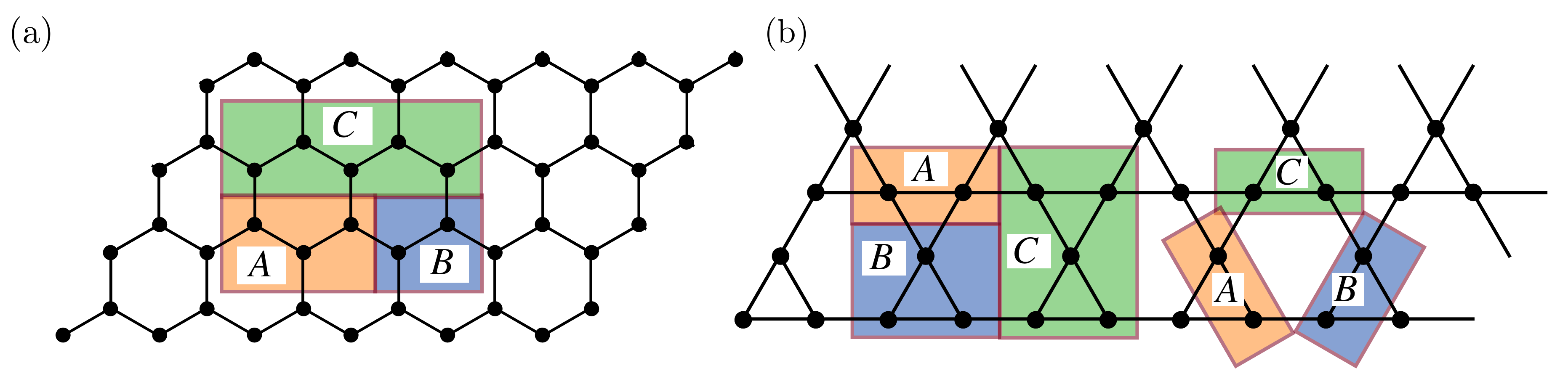}
    \caption{Subsystem partitions used to compute the chiral central charge $c_-$ and the topological entanglement entropy $\gamma$. (\textbf{a}) Kitaev honeycomb model: both $c_-$ and $\gamma$ are obtained from the same tripartition $A,B,C$ of a disk-shaped region. (\textbf{b}) Kagome model: left panel shows the partition used for $c_-$, while the right panel shows the partition used for $\gamma$.}
    \label{fig:Subsystems_c_gamma}
\end{figure}

To evaluate the universal quantities from exact diagonalization, we defined bipartitions of the finite clusters as shown in Fig.~\ref{fig:Subsystems_c_gamma}. For the Kitaev honeycomb model, both the chiral central charge $c_-$ and the topological entanglement entropy (TEE) $\gamma$ were obtained from the same tripartition $A,B,C$ of a disk-shaped region as depicted in Fig.~\ref{fig:Subsystems_c_gamma}a. For the kagome Heisenberg model with scalar spin chirality, the left panel of Fig.~\ref{fig:Subsystems_c_gamma}b shows the tripartition used to evaluate $c_-$, while the right panel shows the partition used to extract $\gamma$. In all cases, the subsystems were chosen sufficiently large compared to the correlation length to reliably capture the universal contributions.

\end{document}